\documentclass[12pt,aps,amsmath,amssymb,prb,reprint,preprintnumbers, nofootinbib, showkeys, showpacs]{revtex4-1}
\usepackage{graphicx, textcomp, wasysym, amssymb, epsfig} 

\begin{document}

\title{Coexistence of magnetic order and valence fluctuations in the Kondo lattice system Ce$_2$Rh$_3$Sn$_5$.}

\author{M. B. Gam\.za$^{1,2,3}$ \email{MGamza@uclan.ac.uk}, R. Gumeniuk$^{2,4}$, U. Burkhardt$^{2}$, W. Schnelle$^{2}$, H. Rosner$^{2}$, A. Leithe-Jasper$^{2}$, and A.~\'{S}lebarski$^{3,5}$}

\affiliation{$^{1}$Jeremiah Horrocks Institute for Mathematics, Physics and Astrophysics, University of Central Lancashire, Preston PR1 2HE, UK}
\affiliation{$^{2}$Max-Planck Institute for Chemical Physics of Solids, D-01187 Dresden, Germany}
\affiliation{$^{3}$Institute of Physics, University of Silesia, 40-007 Katowice, Poland}
\affiliation{$^{4}$Institute of Experimental Physics, TU Bergakademie Freiberg, 09596 Freiberg, Germany}
\affiliation{$^{5}$Centre for Advanced Materials and Smart Structures, Polish Academy of Sciences, 50-950 Wroc\l aw, Poland}

\begin{abstract} 

We report on the electronic band structure, structural, magnetic and thermal properties of Ce$_2$Rh$_3$Sn$_5$. 
Ce~$L_{\mathrm{III}}$--edge~XAS spectra give direct evidence for an intermediate valence behaviour. Thermodynamic measurements reveal magnetic transitions at $T_{\mathrm{N1}}\approx$~2.9~K and $T_{\mathrm{N2}}\approx$~2.4~K. 
Electrical resistivity shows behaviour typical for Kondo lattices.  
The coexistence of magnetic order and valence fluctuations in a Kondo lattice system we attribute to a peculiar crystal structure in which Ce ions occupy two distinct lattice sites. 
Analysis of the structural features of Ce$_2$Rh$_3$Sn$_5$, together with results of electronic band structure calculations, thermodynamic and spectroscopic data indicate that at low temperatures only Ce ions from the Ce1 sublattice adopt a stable trivalent electronic configuration and show local magnetic moments that give rise to the magnetic ordering. By contrast, our study suggests that Ce2 ions exhibit a nonmagnetic Kondo--singlet ground state. Furthermore, the valence of Ce2 ions estimated from the Ce~$L_{\mathrm{III}}$--edge~XAS spectra varies between +3.18 at 6~K and +3.08 at room temperature.  
Thus, our joined experimental and theoretical investigations classify Ce$_2$Rh$_3$Sn$_5$ as a multivalent charge--ordered system.

\end{abstract} 

\pacs{71.28.+d, 75.30.Mb, 71.27.+a, 79.60.-i, 71.20.Lp} 
      
\keywords{Kondo lattice, intermediate valence, mixed valence, electronic structure, magnetic, thermal and transport properties}    

\maketitle

\section{Introduction}

In Ce-based Kondo lattice systems a delicate interplay of localized and itinerant electronic degrees of freedom leads to a wealth of intriguing strongly correlated electron phenomena.\cite{Review1R, SIK, reduced-moments, SC, NFL, Lawrence} 
Kondo lattice compounds contain Ce ions arranged periodically in a crystal lattice. Thus, local $f$--moments of the Ce$^{3+}$ ions can mutually couple via the conduction electrons by means of the Ruderman-Kittel-Kasayu-Yosida (RKKY) interaction.\cite{RKKY} Simultaneously, the $f$--moments are screened by spins of conduction $s$--electrons. This antiferromagnetic (AFM) Kondo coupling drives the demagnetization of the $f$--electron states and leads to the formation of the Abrikosov--Suhl resonance near the Fermi level.\cite{Review1R, SIK} 

Both the RKKY interaction and Kondo effect depend on the coupling constant $J_{s-f}$ of the local $f$--moments with the conduction electron states, with the characteristic temperatures \mbox{$T_{\mathrm{RKKY}}\sim J_{s\mathrm{-}f}^2$} and \mbox{$T_{\mathrm{K}}\sim exp(1$/\ $ |J_{s\mathrm{-}f}|)$,} respectively.\cite{KondoLattice} In the weak coupling limit, the magnetic interaction dominates over the Kondo spin--compensation and a magnetic ground state results.\cite{Review1R, SIK, reduced-moments} For medium $J_{s\mathrm{-}f}$ values, a strong competition between Kondo effect and magnetic interactions gives rise to diverse intriguing physical phenomena, including magnetism with reduced moments,\cite{reduced-moments} non--Fermi liquid behaviour\cite{NFL} or magnetically driven superconductivity.\cite{reduced-moments, SC} In turn, in the strong coupling limit, the Kondo effect predominates and leads to a nonmagnetic heavy--fermion (HF) ground state.\cite{Review1R} Importantly, for large $J_{s\mathrm{-}f}$ values the strong hybridization between the 4$f$ and other conduction band states together with a proximity of the 4$f$ level to the Fermi energy may also trigger instabilities of the charge configuration of the Ce ions, resulting in an intermediate valence (IV) behaviour.\cite{Lawrence} Thus, physical properties of IV systems are governed by both spin fluctuations due to Kondo effect and charge fluctuations between 4$f^0$ and 4$f^1$ configurations that are nearly degenerate in energy.

While vast majority of Ce--based intermetallic compounds contains either trivalent or intermediate--valent Ce ions, only for a limited number of systems a coexistence of both species has been reported. Examples of such materials include Ce$_2$RuZn$_4$\cite{Ce2RuZn4-1, Ce2RuZn4-2}, Ce$_3$Ni$_2$Ge$_7$\cite{Ce3Ni2Ge7}, Ce$_{23}$Ru$_7X_4$ ($X=$ Mg, Cd)\cite{Ce23Ru7X4-1, Ce23Ru7X4-2}, CeRuSn\cite{CeRuSn1, CeRuSn2, CeRuSn3}, Ce$_5$Sn$_3$\cite{8}, Ce$_7T_3$ ($T$--transition metal)\cite{12}, Ce$_{4-x}$Ru$_4$Ge$_{12+x}$\cite{skutterudite1}.
These systems often exhibit remarkable electronic and magnetic properties related to a mixture of Ce ions with long and extraordinarily short distances to the neighbouring atoms due to a peculiar bonding situation. 
Bearing this distinctive structural feature in mind, we started systematic investigations aiming at finding novel intermetallic compounds with highly unconventional magnetic behaviour resulting from the presence of Ce ions in different valence states.

We focused our search on the system Ce--Rh--Sn as it is rich in ternary phases that show a full spectrum of strongly correlated electron phenomena related to various strength of hybridization between Ce~4$f$ and other valence band states.\cite{moje112, PUBL1_112, moje5410, moje124, moje3413, moje3413-new, CeRhSn, CeRhIn} CeRhSn$_2$, Ce$_5$Rh$_4$Sn$_{10}$, CeRh$_2$Sn$_4$ and Ce$_3$Rh$_4$Sn$_{13}$ are magnetically ordered Kondo lattice systems.\cite{moje112, PUBL1_112, moje5410, moje124} In contrast, for Ce$_{3+x}$Rh$_4$Sn$_{13-x}$ \mbox{($0.2 \apprle x \apprle 0.6$)} no sign of Kondo effect or long range magnetic order was found even down to 0.4~K.\cite{moje3413-new} In turn, in CeRhSn the Ce ions are in an IV state.\cite{CeRhSn, CeRhIn}

Here, we report on Ce$_2$Rh$_3$Sn$_5$ that crystallizes in the orthorhombic Y$_2$Rh$_3$Sn$_{5}$ type of structure, where Y ions occupy two distinct crystallographic sites.\cite{Patil235, R5} 
An early study revealed a moderate HF behaviour \mbox{($\gamma \approx$ 150 mJ Ce-mol$^{-1}$K$^{-2}$)} with a magnetic transition at \mbox{T$_{\mathrm{N}} \approx$ 5 K} (from resistivity), 4~K (from magnetic susceptibility) or 2.5~K (from heat capacity).\cite{Patil235} Our single crystal X-ray diffraction (XRD) study unveils extraordinarily short Ce--Rh contacts indicative of valence larger than 3+ for one Ce site. Interestingly, the local environment of Ce atoms from the second sublattice is very similar to that in CeRh$_2$Sn$_4$, a magnetic Kondo lattice system with trivalent Ce ions and \mbox{T$_{\mathrm{N}}\approx 3.2$ K.\cite{moje124}}
Motivated by these results, we performed a combined experimental and theoretical study on Ce$_2$Rh$_3$Sn$_5$ based on thermodynamic measurements and spectroscopic data together with first principles electronic structure calculations aiming at exploring its complex structural and magnetic properties.

\section{Methods}

\subsection{Experimental}

Polycrystalline samples of Ce$_2$Rh$_3$Sn$_5$ with total weight of about 2~g were prepared from ingots of cerium (Ames, 99.9~wt.\%), rhodium granules (ChemPur, 99.9~wt.\%) and tin foil (ChemPur, 99.995~wt.\%). Stoichiometric amounts of the elemental metals were arc-melted on a water cooled copper hearth in an ultra-high purity Ar atmosphere with a Zr getter (heated above the melting point). The sample was remelted several times to promote homogeneity and heat-treated at 800$^{\circ}$C for 14 days in a sealed Ta tube enclosed in an evacuated quartz tube. Almost no mass loss (below~0.02\%) occurred during the melting and annealing processes. 

All manipulations related to preparation of the sample and its storage were performed in a argon--filled glove box (MBRAUN, \mbox{p(O$_2$/H$_2$O) $\leq$ 1 ppm)} in order to prevent the oxidation.

The quality of the sample was examined by means of powder XRD measurements and metallographic investigations. Details of these studies are included in the Supplementary Information\cite{Supplementary}. Powder XRD pattern indicates that the sample is nearly single phased. Microprobe measurements revealed the chemical composition that corresponds to Ce$_{2.01(1)}$Rh$_{3.00(2)}$Sn$_{5.00(2)}$ and thus confirms the desired stoichiometry. 

An irregularly shaped crystal was mechanically extracted from the annealed ingot. Single crystal XRD study was performed at room temperature using an Xcalibur~E Single Crystal Diffractometer. Details concerning data collection and handling are summarized in Table~\ref{tab:Table1}. Structure refinements were carried out using the Jana2006 program.~\cite{jana2006}

The \mbox{Ce $L_{\mathrm{III}}$ XAS} spectra were recorded in transmission arrangement at the EXAFS beamline~C of HASYLAB at DESY at the temperatures of 80~K and 293~K. The wavelength selection was realized using the four-crystal mode which yielded an experimental resolution of $\sim$2~eV (FWHM) at the Ce~$L_{\mathrm{III}}$ threshold of 5723~eV. Powdered samples of Ce$_2$Rh$_3$Sn$_5$ were mixed with small amounts of B$_4$C and mounted on 1~cm$^2$ window sample holders using paraffin wax. Two series of measurements performed using different sample powders at ambient temperature and at low temperatures down to 6~K using a helium gas flow cryostat gave consistent results. Experimental data were recorded with CePO$_4$ as the external reference compound with Ce$^{3+}$ ions. The Ce~$L_{\mathrm{III}}$~XAS spectra were evaluated using the Athena program package\cite{Athena}. 

XPS experiments were performed at room temperature using a PHI~5700~ESCA spectrometer with monochromatized Al~K$_{\alpha}$ radiation. The energy resolution was about 0.4~eV. The polycrystalline sample was broken under a high vacuum of $6\times 10^{-10}$~Torr immediately before measuring the spectra. Binding energies were referenced to the Fermi level \mbox{($E_{\mathrm{F}}=0$).} Calibration of the spectra was performed according to Ref.~\onlinecite{ZZ}.

The magnetization studies were carried out in a SQUID magnetometer (MPMS~XL-7, Quantum Design) at temperatures between 1.8~K and 400~K in magnetic fields up to 70~kOe. Electrical resistivity measurements were performed with a standard dc four--probe set--up. Heat capacity was determined by a relaxation--type  method using a commercial system (PPMS, Quantum  Design).   

\begin{table}
\caption{\label{tab:Table1} Crystallographic data for Ce$_2$Rh$_3$Sn$_5$.}
\begin{tabular}{lll}
 \hline
Empirical formula   & & Ce$_2$Rh$_3$Sn$_5$ \\
Structure type      & &Y$_2$Rh$_3$Sn$_5$ \\
Space group         & &$Cmc2_1$ (No. 36) \\
Lattice parameters$^a$ & & $a$ = 4.4992(1) \r{A} \\
                         & & $b$ = 26.4839(7)  \r{A} \\
                         & & $c$ = 7.2160(2)  \r{A} \\
Unit cell volume,$^a$ $V$ & & 859.83(4) \r{A}$^3$ \\
Formula units/cell, $Z$ & & 4 \\
Crystal density, $\rho$ & & 9.151 g cm$^{-3}$ \\ 
Temperature & & 295(5) K \\ 
Diffraction system & & Xcalibur E, four-circle Kappa \\
                   & & Sapphire CCD Detector (Xcalibur)\\
Radiation, $\lambda$  & & Mo $K\alpha$, 0.71073 \r{A} \\
Range in $h,k,l$  & & $\pm$9, $\pm$52, $\pm$14 \\
$R(eqv)$/$R(\sigma)$ & & 0.078/0.018 \\
2$\theta_{min}$/2$\theta_{max}$  & & 4.17/45.49 \\
Observation criteria & & $F$($hkl$) \textgreater 3.00$\sigma(F)$ \\
Resolution d (\r{A}) & & 0.45 \\
Absorption correction & & face--based, analytical\cite{CR} \\
Absorption coefficient & & 30.194 \\
N($hkl$) measured & & 71975  \\
N($hkl$) unique    & & 3939  \\
Extinction method & &  isotropic type 2 correction\cite{Becker-and-Coppens-1974}  \\
Extinction coefficient & &    4070(70)  \\
Number of parameters & & 63 \\
Goodness-of-fit (GOF)	& & 1.94    \\
R                 & & 2.25\%    \\ 
wR                & & 3.28\%    \\ \hline
\multicolumn{3}{l}{$^a$powder data} \\
 \end{tabular}
\end{table}

\subsection{Computational} 
\label{computational}

The electronic structure of Ce$_2$Rh$_3$Sn$_5$ was studied using the Full Potential Local Orbital (FPLO) Minimum Basis code\cite{FPLO}. Scalar--relativistic calculations based on Density Functional Theory (DFT) were performed within the local (spin) density approximation [L(S)DA] using the Perdew and Wang\cite{LSDAPW} parametrization of the exchange--correlation (XC) potential. The strong Coulomb correlation within the Ce~4$f$ shell was also treated in a mean field approximation using the LSDA+$U$ method\cite{LSDAU} with the around mean field double counting scheme. The Coulomb repulsion $U$ and exchange constant $J$ for the 4$f$ states of both types of Ce atoms were assumed to be 1--8~eV and 1~eV, respectively. Thus, the effective \mbox{$U_{\mathrm{eff}}$=$U$-$J$} was from the range of \mbox{0--7~eV.}
The Brillouin zone sampling was based on 198 {\bf k}--points in the irreducible wedge (2000 points in the full zone). A series of calculations with an increasing density of the {\bf k}--mesh was performed to ensure the convergence of the total energy with respect to the {\bf k}--grid.   

Based on the band structure results we estimated the theoretical XPS valence band spectra. The partial $l$--resolved densities of states obtained using the LDA and the LSDA+U ($U$=6~eV, $J$=1~eV) methods were weighted by their respective photoionisation cross--sections\cite{cros}. The results were multiplied by the Fermi--Dirac function for 300~K and convoluted by Lorentzians with a full width at half maximum (FWHM) of 0.4~eV to account for the instrumental resolution, thermal broadening and the lifetime effect of the hole states.

\section{Results and discussion}

\subsection{Crystal structure} 
\label{structure}

\begin{figure}
\includegraphics[width=0.45\textwidth,angle=0]{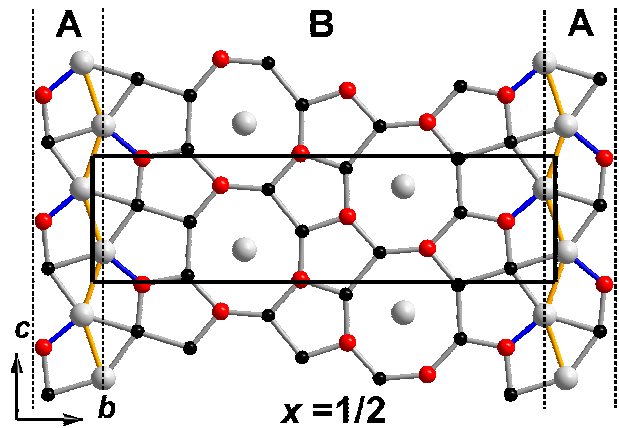}
\caption{\label{fig:Fig1} (Color online) The planar layer in the structure of Ce$_2$Rh$_3$Sn$_5$ occurring at $x = 1/2$. Fragment A: a chain of interconnected Ce atoms (yellow line) and edge sharing tetragons and triangles; fragment B: condensed empty pentagons and octagons of Rh (red) and Sn (black) centred by Ce (grey) atoms.}  
\end{figure}

\begin{table*}
\caption{\label{tab:Table2} Atomic positional and displacement parameters for Ce$_2$Rh$_3$Sn$_5$ (Note: $B_{12}$=$B_{23}$=0.) All atoms are located at 4$a$ (0,y,z) Wyckoff positions. The experimental structural data are compared to those derived from the band structure calculations in the LDA approximation. The free parameters in atomic coordinates obtained from the computational study were rounded to three significant digits.}
\begin{tabular}{l|l|l|l|l|l|r@{.}l|l|l|l}
     &    \multicolumn{8}{|c|}{experiment}  &  \multicolumn{2}{c}{LDA}  \\ \cline{2-11}
Atom &  $y$       & $z$          & $B_{11}$ & $B_{22}$ & $B_{33}$ & \multicolumn{2}{l}{$B_{23}$} & $B_{iso}$  &  $y$  & $z$   \\ \hline  
Ce1  & 0.32754(1) & 0.27397(5)   & 0.72(1)  & 0.77(1)  & 0.98(1)  &-0&02(1) 					 & 0.82(1)    & 0.330 & 0.274 \\
Ce2  & 0.02426(1) & 0.25076(4)   & 0.93(1)  & 0.77(1)  & 0.81(1)  &-0&08(1) 					 & 0.84(1)    & 0.026 & 0.239 \\
Rh1  & 0.72296(2) & 0.28274(6)   & 0.75(1)  & 0.74(1)  & 0.87(2)  &-0&06(1) 					 & 0.79(1)    & 0.723 & 0.280 \\
Rh2  & 0.10594(2) & 0.00000(6)   & 1.06(2)  & 0.82(1)  & 0.94(2)  & 0&01(1) 					 & 0.94(1)    & 0.103 & 0.999 \\
Rh3  & 0.44897(2) & 0.03180(6)    & 0.78(1)  & 0.82(1)  & 0.80(2)  & 0&00(1) 					 & 0.80(1)    & 0.450 & 0.015 \\
Sn1  & 0.21037(1) & 0.49104(5)    & 0.80(1)  & 0.82(1)  & 0.73(1)  &-0&07(1) 					 & 0.78(1)    & 0.210 & 0.491 \\
Sn2  & 0.20563(1) & 0.07177(6)    & 0.71(1)  & 0.82(1)  & 0.79(1)  & 0&09(1) 					 & 0.77(1)    & 0.205 & 0.065 \\
Sn3  & 0.62182(1) & 0.25218(6)    & 1.41(1)  & 0.69(1)  & 0.93(1)  &-0&09(1) 					 & 1.01(1)    & 0.622 & 0.242 \\
Sn4  & 0.45116(1) & 0.41543(6)    & 0.86(1)  & 0.73(1)  & 0.77(1)  & 0&01(1) 					 & 0.79(1)    & 0.452 & 0.400 \\ 
Sn5  & 0.09635(1) & 0.62444(6)    & 0.77(1)  & 0.76(1)  & 0.99(1)  & 0&16(1) 					 & 0.84(1)    & 0.096 & 0.618 \\ \hline
 \end{tabular}
\end{table*}

\begin{table}
\caption{\label{tab:Table3} Selected interatomic distances in Ce$_2$Rh$_3$Sn$_5$.}
\begin{tabular}{c@{ }ll| c@{ }ll}
\hline
\multicolumn{2}{l}{Atoms}  & $d$ (\r{A}) & \multicolumn{2}{l}{Atoms}  & $d$ (\r{A}) \\ \hline
Ce1 --& 2Sn1& 3.1979(4)    &  	   &  1Ce1& 3.4733(5)   \\
      & 2Sn5& 3.2060(4)    &       &  1Sn2& 3.0255(5)   \\ 
      & 2Sn2& 3.2310(4)    &       &  1Sn5& 3.1670(5)   \\ 
      & 1Sn4& 3.4268(5)    &       &  2Sn2& 3.2155(4)   \\     
      & 1Sn1& 3.4733(5)    &       &  2Rh1& 2.7250(3)   \\
      & 1Sn2& 3.5402(5)    &       &  1Rh1& 2.7452(6)   \\ \cline{4-6} 
      & 2Rh2& 3.2885(4)    & Sn2 --&  2Ce1& 3.2310(4)    \\ \cline{1-3}      
Ce2 --&  2Sn4& 3.1954(3)   & 	   &  1Ce1& 3.5402(5)    \\
      &  1Sn5& 3.3012(5)   &       &  1Sn1& 3.0255(5)   \\
      &  1Sn5& 3.3190(5)   &       &  2Sn1& 3.2155(4)    \\ 
      &  2Sn4& 3.3656(4)   &       &  1Rh2& 2.6885(6)  \\ 
      &  2Sn3& 3.4243(4)   &       &  2Rh1& 2.7537(4)   \\
      &  1Rh2& 2.8180(5)   &       &  1Rh1& 2.8132(6)    \\ \cline{4-6}  
      &  2Rh3& 3.1091(4)   & Sn3 --&  2Ce2& 3.4243(4)   \\ \cline{1-3}       
Rh1 --&  1Sn3& 2.6854(5)   &	   &  1Sn4& 3.1024(5)   \\
      &  2Sn1& 2.7250(3)   &       &  1Rh1& 2.6854(5)   \\
      &  1Sn1& 2.7452(6)   &       &  1Rh3& 2.7522(6)   \\
      &  2Sn2& 2.7537(4)   &       &  2Rh2& 2.9226(4)   \\ \cline{4-6}
      &  1Sn2& 2.8132(6)   & Sn4 --&  2Ce2& 3.1954(3)   \\ \cline{1-3}      
Rh2 --&  1Ce2& 2.8180(5)   &	   &  2Ce2& 3.3656(4)   \\
      &  2Ce1& 3.2885(4)   &       &  1Ce1& 3.4268(5)   \\
      &  1Sn2& 2.6885(6)   &       &  1Sn3& 3.1024(5)   \\
      &  1Sn5& 2.7196(6)   &       &  1Rh3& 2.7665(6)   \\
      &  2Sn4& 2.7774(3)   &       &  1Rh3& 2.7729(6)   \\
      &  2Sn3& 2.9226(4)   &       &  2Rh2& 2.7774(3)   \\ \cline{1-6}
Rh3 --&  2Ce2& 3.1091(4)   & Sn5 --&  2Ce1& 3.2060(4)   \\      
      &  2Sn5& 2.6350(3)   &       &  1Ce2& 3.3012(5)   \\
      &  1Sn3& 2.7522(6)   &       &  1Ce2& 3.3190(5)   \\
      &  1Sn4& 2.7665(6)   &       &  1Sn1& 3.1670(5)    \\
      &  1Sn4& 2.7729(6)   &       &  2Rh3& 2.6350(3)   \\ \cline{1-3}
Sn1 --&  2Ce1& 3.1979(4)   &       &  1Rh2& 2.7196(6)   \\ \hline  
 \end{tabular}
\end{table}

Refinement of the single crystal XRD data shows that Ce$_2$Rh$_3$Sn$_5$ crystallizes with the non-centrosymmetric orthorhombic Y$_2$Rh$_3$Sn$_5$ type of structure (space group $Cmc2_1$), in agreement with earlier reports.\cite{Patil235, R5} The crystallographic details of the refinement are given in Table~\ref{tab:Table1}, atomic coordinates and anisotropic displacement parameters -- in Table~\ref{tab:Table2} and interatomic distances in Table~\ref{tab:Table3}.

The Y$_2$Rh$_3$Sn$_5$ type is a layered structure consisting of two analogical layers shifted by 1/2 of the translation in the $bc$-plane and alternating along $x$-direction. As indicated in Fig.~\ref{fig:Fig1}, each layer consists of two main fragments: A (a chain of interconnected Y2/Ce2 atoms and edge sharing tetragons and triangles) and B (condensed pentagons and octagons). The octagons are centred by heavy Y1/Ce1 atoms, while pentagons remains empty. The B-fragments are very similar to layers formed by octagons and pentagons in the structure of the NdRh$_2$Sn$_4$ type. The close structural relationship of Y$_2$Rh$_3$Sn$_5$ and NdRh$_2$Sn$_4$ types is widely discussed in the literature.\cite{moje124, R5, R6}

The interatomic distances in the structure of Ce$_2$Rh$_3$Sn$_5$ (Table~\ref{tab:Table3}) mostly correlate well with the sum of atomic radii of the elements [$r_{\mathrm{(Ce)}}=1.825$~\r{A}; $r_{\mathrm{(Rh)}}=1.34$~\r{A}; $r_{\mathrm{(Sn)}}=1.41$~\r{A}]\cite{R4}. The Sn--Sn contacts are slightly longer than 2.81~\r{A}, while the Rh--Sn distances are shortened by about 4-5\%, which assumes the formation of covalently bonded Rh--Sn framework in the investigated structure. The shrinking of Ce--Sn contacts is of 1-2\%, assuming also a weak interaction.

The most intriguing feature of the Ce$_2$Rh$_3$Sn$_5$ structure is the shortened Ce2--Rh2 distance by nearly 11\% as compared to the sum of atomic radii of Ce and Rh. 
Such extraordinarily short Ce2--Rh contacts indicate that the valence of Ce2 ions should be larger than +3.\cite{CeRhIn, Lawrence, Fug83, 235Kaczorowski} 
In turn, for Ce1 ions all the Ce1--Sn and Ce1--Rh distances are nearly equal to the sum of corresponding atomic radii. This suggests that the electronic configuration for Ce1 ions is close to 4$f^1$, i.e. Ce$^{+3}$. 
Thus, the analysis of the crystal structure indicates that Ce$_2$Rh$_3$Sn$_5$ may be a mixed valence, IV or even a multivalent charge--ordered system. The last scenario assumes a static ordering of trivalent and intermediate--valent Ce ions in two distinct lattice sites and was proposed for systems such as Ce$_2$RuZn$_4$\cite{Ce2RuZn4-1, Ce2RuZn4-2}, Ce$_3$Ni$_2$Ge$_7$\cite{Ce3Ni2Ge7}, Ce$_{23}$Ru$_7X_4$ ($X=$ Mg, Cd)\cite{Ce23Ru7X4-1, Ce23Ru7X4-2}, Ce$_5$Sn$_3$\cite{8}, YbPtGe$_2$.\cite{gumeniuk2012_Yb}.

To inspect the valence states of Ce ions in Ce$_2$Rh$_3$Sn$_5$ more closely, we performed spectroscopic measurements.

\subsection{Ce $L_{\mathrm{III}}$ XAS}
\label{XAS-spectra}

\begin{figure}
\includegraphics[width=0.44\textwidth,angle=0]{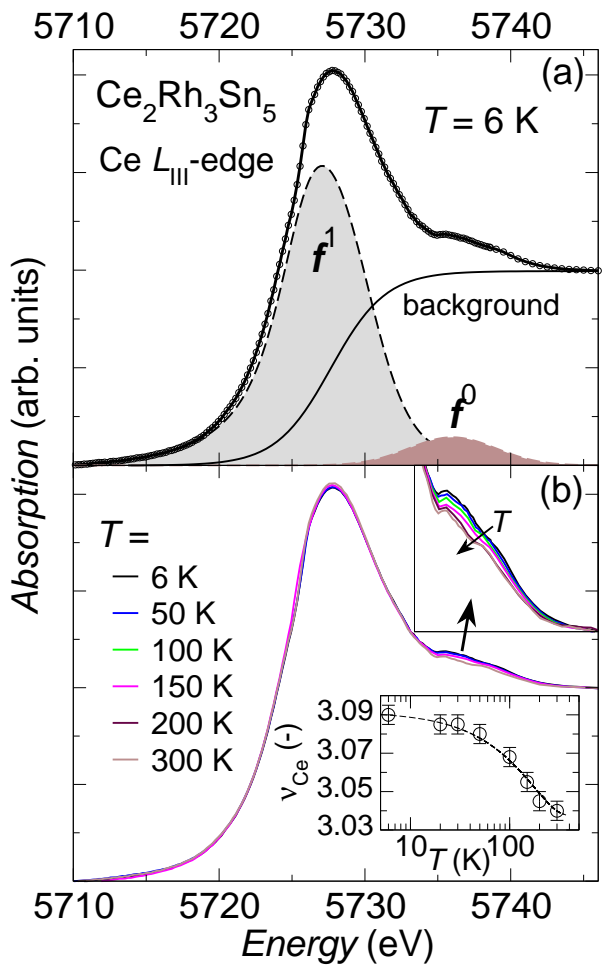}
\caption{\label{fig:Fig2} (Color online) Near--edge regime of Ce $L_{\mathrm{III}}$ XAS spectra for Ce$_2$Rh$_3$Sn$_5$. Panel a shows deconvolution of the normalized spectrum recorded at $T=6$~K. The contributions due to $f^1$ and $f^0$ configurations are indicated by beige or grey fields underneath the dashed lines, while the solid black line represents the arctan function that accounts for transitions of 2$p_{\mathrm{3/2}}$ electrons to the extended conduction states. 
Panel b depicts normalized Ce $L_{\mathrm{III}}$ XAS spectra measured at various temperatures between 6~K and 300~K. The changes in intensity of the $f^0$ contribution are enlarged in the upper inset, with the arrow indicating the direction of increasing temperature. Lower inset of panel b displays calculated changes of Ce valence as a function of temperature. The dashed line was guided to the eye. 
}
\end{figure}

Fig.~\ref{fig:Fig2} presents the near--edge regime of Ce~$L_{\mathrm{III}}$~XAS spectra for Ce$_2$Rh$_3$Sn$_5$ that were recorded at several temperatures between 6~K and 300~K. Although the spectra are dominated by 'white line' at the energy of $\sim$5726~eV corresponding to 4$f^1$ configuration of Ce$^{3+}$, there is also an additional contribution at the energy of nearly +9~eV above the 'white line' maximum that indicates the presence of Ce$^{4+}$ species with electron configuration 4$f^0$. 
Increasing temperature from 6~K to 300~K leads to a gradual reduction of the high energy contribution due to 4$f^0$ configuration of Ce$^{4+}$ and a simultaneous slight increase in intensity of the 'white line', as shown in Fig.~\ref{fig:Fig2}b. Such a progressive shift of the spectral weight to lower energies implies an IV behaviour. 
Although the temperature induced changes in the Ce~$L_{\mathrm{III}}$~XAS spectra are rather small, similar slight effects were reported for many Ce--based IV compounds.\cite{235Kaczorowski,IVXAS3, IVXAS4, Lawrence}  

Deconvolution of the Ce~$L_{\mathrm{III}}$~XAS spectra (example in Fig.~\ref{fig:Fig2}a) indicates that the mean Ce valence in Ce$_2$Rh$_3$Sn$_5$ decreases from +3.09(1) at 6~K to +3.04(1) at ambient temperature, with the most rapid changes taking place at temperatures of 100~K (see lower inset in Fig.~\ref{fig:Fig2}b).
However, since the measured spectra contain signal originating from all Ce ions, the obtained values of Ce valence in Ce$_2$Rh$_3$Sn$_5$ should be interpreted as the $average$ valence for Ce ions from two distinct lattice sites. Therefore, assuming that Ce1 ions are trivalent in the entire temperature range, the valence of Ce2 species should vary between +3.18 at 6~K and +3.08 at $\sim$300~K. We note that similar Ce valence was reported for nonmagnetic IV compounds such as CeRhSi$_2$\cite{235Kaczorowski}, Ce$_2$Ni$_3$Si$_5$\cite{235CeNiSi}, CeMo$_2$Si$_2$C\cite{CeMo2Si2C}. 

To sheed more light on the character of Ce~4$f$ states in Ce$_2$Rh$_3$Sn$_5$, we performed X--ray photoelectron spectroscopy measurements.

\subsection{Ce 3$d$ XPS}
\label{core}

\begin{figure}
\includegraphics[width=0.46\textwidth,angle=0]{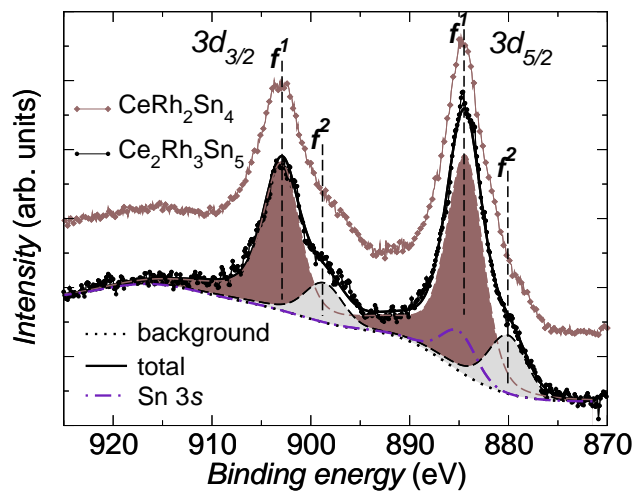} 
\caption{\label{fig:Fig3} (Color online) The Ce 3$d$ XPS spectra for Ce$_2$Rh$_3$Sn$_5$ and CeRh$_2$Sn$_4$. The spin--orbit components 3$d_{3\/2}$ and 3$d_{5\/2}$ as well as the $f^1$ and $f^2$ contributions are labelled. The deconvolution of the spectrum for Ce$_2$Rh$_3$Sn$_5$ is shown.} 
\end{figure}

Fig.~\ref{fig:Fig3} shows room temperature Ce~3$d$~XPS spectrum for Ce$_2$Rh$_3$Sn$_5$ together with the data for the structurally related compound CeRh$_2$Sn$_4$. Due to the spin--orbit (SO) coupling, Ce 3$d$ states give rise to two sets of photoemission lines that correspond to the $j=3/2$ and $j=5/2$ components of final states. Each of these SO sets consists of two distinct contributions labelled as $f^1$ and $f^2$. The $f^1$ components originate from the screening of the core hole by conduction electrons. The $f^2$ satellites arise when an electron is transferred from extended valence band states to the 4$f$ states during photoemission process in order to screen the hole in the core shell. Since the probability of such processes depends on the coupling of the 4$f$ levels to the other states near the Fermi level,\cite{Fug83} the presence of the pronounced $f^2$ peaks indicates that there is a notable hybridization between the Ce 4$f$ and conduction band states in Ce$_2$Rh$_3$Sn$_5$. 

There are no additional peaks in the Ce~3$d$~XPS spectrum of Ce$_2$Rh$_3$Sn$_5$ in a distance of $\sim$11~eV from the main photoemission lines, that could be assigned to the 4$f^0$ final states. This result may seem to contradict the mean Ce valence of +3.04 at ambient temperature obtained from our XAS measurements (see Section~\ref{XAS-spectra}). We note, however, XPS measurements with energy resolution of 0.4~eV may not show weak $f^0$ contributions. 
Our simulations (not shown) indicate that only peaks with intensities greater than $\sim$5\% of the combined intensity of $f^1$ and $f^2$ contributions should be detectable in the Ce~3$d$~XPS spectrum of Ce$_2$Rh$_3$Sn$_5$, whereas weaker $f^0$ contributions should be hidden in the background signal and experimental noise.  
Thus, the absence of $f^0$--type peaks implies the mean occupancy of the Ce~4$f$ shell \mbox{$n_f\gtrsim 0.95$} at room temperature, which agrees with the results of our XAS study (see Section~\ref{XAS-spectra}).

Deconvolution of the Ce~3$d$~XPS spectrum was performed on the basis of Doniach-\v Sunji\'c theory\cite{Don70}. The intensity ratio \mbox{$I(3d_{5/2}) / I(3d_{3/2}) = 3/2$} was fixed during the fitting, as required by the quantum--mechanical rules. The SO split \mbox{$\delta_{\mathrm{S-O}}$ $\approx$ $18.6$ eV} was assumed. A Tougaard--type background\cite{Tou} was subtracted from the XPS data. A small peak due to the Sn~3$s$ states located at the binding energy of 885~eV was included in the fit, with the intensity determined by the stoichiometry of the compound Ce$_2$Rh$_3$Sn$_5$. Model calculations of Gunnarsson and Sch\"onhammer (GS)\cite{Fug83, Gun83} were used to calculate the value of the $\Delta$ parameter describing the hybridization strength between the Ce~4$f$ shell and conduction electron states from relative intensities of the $f^2$ peaks. Such a procedure yielded \mbox{$\Delta \approx$ 100 meV,} which should be considered as the $average$ hybridization parameter for Ce ions occupying two lattice sites in Ce$_2$Rh$_3$Sn$_5$. 

The obtained $\Delta$ value is comparable to those found in systems with Ce in an IV state \mbox{($\Delta \gtrsim 100$ meV)\cite{CeRhSn, CeRhIn, Fug83}} and is notably stronger than in other compounds from the ternary Ce--Rh--Sn system with trivalent Ce ions \mbox{($\Delta \lesssim 80$ meV)\cite{moje112, moje5410, moje124, moje3413, moje3413-new}.} 
In particular, the $\Delta$ value for Ce$_2$Rh$_3$Sn$_5$ is larger than that for the structurally related compound CeRh$_2$Sn$_4$ with Ce$^{3+}$, as evidenced directly by larger intensities of $f^{\mathrm{2}}$ satellites in the Ce~3$d$~XPS spectra (see Fig.~\ref{fig:Fig3}).

Despite the strong hybridization $\Delta$ and valence fluctuations revealed by our spectroscopic investigations, thermodynamic measurements indicate a magnetic ground state for Ce$_2$Rh$_3$Sn$_5$.

\subsection{Magnetic measurements}
\label{susceptibility}

\begin{figure*}
\includegraphics[width=0.95\textwidth,angle=0]{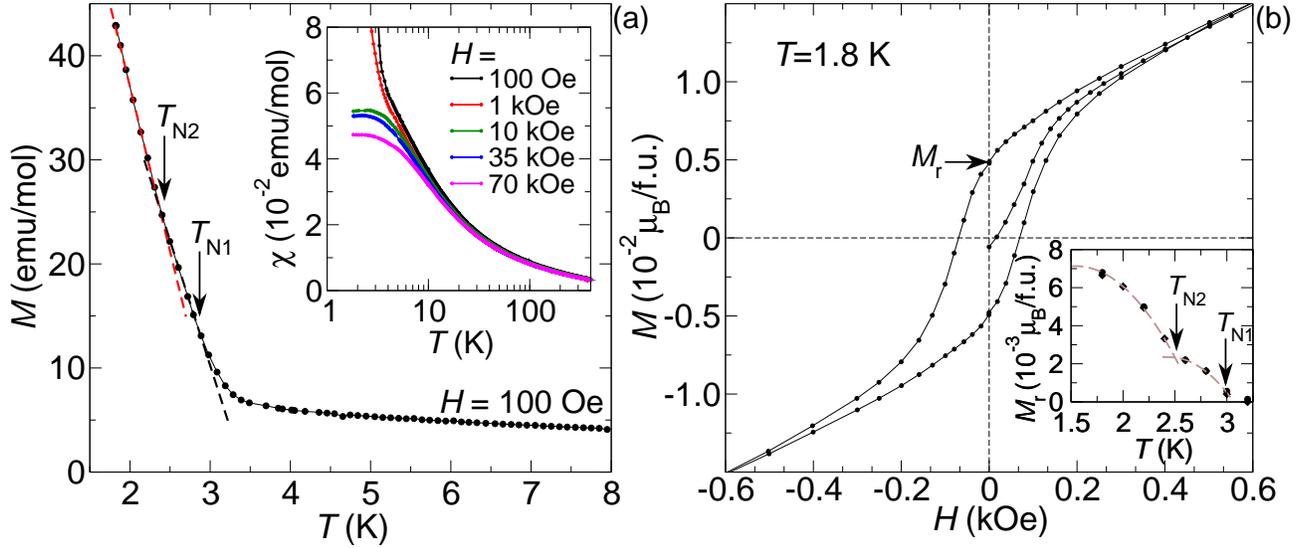}
\caption{\label{fig:Fig6} (Color online) Results of $dc$ magnetization measurements for Ce$_2$Rh$_3$Sn$_5$. Panel a shows the low temperature part of the $M$($T$) measured in magnetic field of 100~Oe. Two distinct changes in slope of the $M$($T$) curve suggestive of magnetic transitions are indicated. The inset displays the magnetic susceptibility measured in a number of magnetic fields between 100~Oe and 70~kOe and plotted as a function of temperature on a logarithmic scale. Panel b depicts the full magnetization loop at $T=1.8$~K. Inset shows values of the remanence magnetization at various temperatures. The dashed lines are guided to the eye.}
\end{figure*}

\begin{figure}
\includegraphics[width=0.46\textwidth,angle=0]{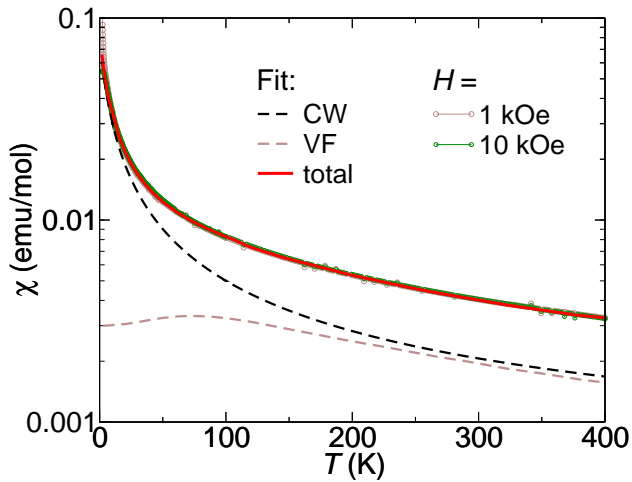}
\caption{\label{fig:Fig7} (Color online) The $dc$ magnetic susceptibility of Ce$_2$Rh$_3$Sn$_5$ measured in magnetic fields of 1~kOe (brown circles) and 10~kOe (green circles), together with the fit using Eq.~\ref{eq:chi-fit} (thick solid red line). The Curie--Weiss--type contribution $\chi_{\mathrm{CW}}$ due to Ce1 sublattice and the $\chi_{\mathrm{Ce2}}$ term estimated based on calculations of Rajan\cite{Rajan} to account for the magnetic susceptibility originating from Ce2 ions are indicated with black and brown dashed lines, respectively. The solid blue line represents the best fit to the $\chi$($T$) data at temperatures of 8--85~K by using the modified Curie--Weiss law.  The inset shows $\chi^{-1}$($T$) in a magnetic field of 10~kOe (green circles) together with the fit by using the modified Curie-–Weiss law that covers the data above $\sim$170~K (dashed red line).
}
\end{figure}

Figures~\ref{fig:Fig6} and \ref{fig:Fig7} present results of a $dc$ magnetization study on polycrystals of Ce$_2$Rh$_3$Sn$_5$. 
The $M$($T$) measured in weak magnetic fields increases rapidly at temperatures below $\sim$2.9~K, suggestive of an onset of magnetic order (see Fig.~\ref{fig:Fig6}a). 
Closer inspection shows that the $M$($T$) grows even faster at \mbox{$T\leq2.4$~K}. This points to a change in magnetic structure at 2.4~K, in agreement with our specific heat data that revealed two anomalies at $T_{\mathrm{N1}}\approx$~2.9~K and $T_{\mathrm{N2}}\approx$~2.4~K (see Section~\ref{specific-heat}). The presence of two magnetic transitions in Ce$_2$Rh$_3$Sn$_5$ provides an explanation for the discrepancy between ordering temperatures reported by Patil $et$ $al$.\cite{Patil235} based on the specific heat and magnetization data.  

Isothermal magnetization curves measured at temperatures below 3~K show distinct hysteresis between the data collected with increasing and decreasing field strengths. The illustrative $M$($H$) data recorded at $T=1.8$~K are shown in Fig.~\ref{fig:Fig6}b. Both the remanence magnetization $M_r$ and the coercive field decrease gradually with increasing temperature and finally diminish at $T_{\mathrm{N1}}\approx2.9$~K (see the lower inset of Fig.~\ref{fig:Fig6}b). However, the magnetization in the ordered states is very small, about two orders of magnitude smaller than magnetization values expected for a ferromagnetic order that involves one magnetic ion per formula unit with $S=1/2$. Thus, the magnetization measurements suggest an AFM ordering with only small canting in Ce$_2$Rh$_3$Sn$_5$ at temperatures below 2.4~K, with the canting becoming even smaller between 2.4~K and 2.9~K.

At high temperatures the magnetic susceptibility $\chi$=$M$/$H$ does not depend on the applied magnetic field and resembles behaviour expected for a local moment paramagnet (see the inset of Fig.~\ref{fig:Fig6}). 
As shown in the inset of Fig.~\ref{fig:Fig7}, at temperatures between $\sim$170~K and 400~K the $\chi^{-1}$($T$) follows a modified Curie–-Weiss law:
\begin{equation}
\chi(T)= \frac{C_{\mathrm{HT}}}{T - \theta_{\mathrm{HT}}} + \chi_{\mathrm{0-HT}},
\label{eq:chi-CW}
\end{equation}
with \mbox{$\chi_{\mathrm{0-HT}}\approx 8\times10^{-5}$~emu/mol}, the paramagnetic Weiss temperature $\theta_{\mathrm{HT}}\approx-103$~K and the Curie constant $C_{\mathrm{HT}}$~=~1.60~emu~K/mol. The obtained $C_{\mathrm{HT}}$ value corresponds to the fluctuating magnetic moment of 2.51~$\mu_\mathrm{B}$ per Ce, which is very close to the effective moment of 2.53~$\mu_\mathrm{B}$/Ce expected for free Ce$^{3+}$ ions. Thus, the magnetic measurements indicate that at temperatures above $\sim$170~K the valence of Ce ions in Ce$_2$Rh$_3$Sn$_5$ is close to 3+ and all the Ce ions bear localized magnetic moments. 

The small positive $\chi_{\mathrm{0-HT}}$ is in--line with metallic properties of Ce$_2$Rh$_3$Sn$_5$. The $\chi_{\mathrm{0-HT}}$ is the sum of the diamagnetic susceptibility of the closed--shells and the conduction electron contributions. Assuming that for tin $\chi_\mathrm{dia}$(Sn$^{4+}$)~=~$-16\times10^{-6}$~emu/mol, for cerium  \mbox{$\chi_\mathrm{dia}$(Ce$^{3+}$)~=} $-20\times10^{-6}$~emu/mol and for rhodium  \mbox{$\chi_\mathrm{dia}$(Rh$^{4+}$)~=~$-18\times10^{-6}$~emu/mol,\cite{increm}} the sum of the diamagnetic core increments yields $-174\times10^{-6}$~emu/mol. Thus, the rough estimate of the electronic Pauli susceptibility $\chi_{\mathrm{P}}$, after correcting for the Landau electron diamagnetism \mbox{$\chi_{\mathrm{L}}$=-$\frac{1}{3}$ $\chi_{\mathrm{P}}$} and the core--level diamagnetism, gives $381\times10^{-6}$~emu/mol. This value corresponds to the DOS($E_{\mathrm{F}}$)~$\approx$~11.8~states~eV$^{-1}$~f.u.$^{-1}$ which is comparable to the value of the DOS($E_{\mathrm{F}}$) obtained from our electronic band structure calculations (Fig.~\ref{fig:Fig9}).

At lower temperatures, the experimental $\chi$($T$) data can be well described by the modified Curie–-Weiss law (see Fig.~\ref{fig:Fig7}) with a small negative Weiss temperature \mbox{$\theta_{\mathrm{LT}}\approx-6$~K} of the order of $T_{\mathrm{N1}}$,  \mbox{$\chi_{\mathrm{LT}}\approx 0.0039$~emu/mol} and the Curie constant $C_{\mathrm{LT}}$~=~0.48~emu~K/mol, which is strongly reduced as compared to the value of 1.60~emu~K/mol expected for full moments of Ce$^{3+}$ ions.   
The hefty lowering of the effective fluctuating moment accompanied by a strong enhancement of the Pauli susceptibility upon lowering temperature points to the delocalization of 4$f$ electrons from some of Ce ions in Ce$_2$Rh$_3$Sn$_5$. The large negative Weiss temperature \mbox{$\theta_{\mathrm{HT}} \approx -110$~K} indicates that there is a strong AFM coupling of the local 4$f$ moments with conduction band states. 
In Ce--based Kondo lattice systems, this interaction drives the demagnetization of the $f$--electron states and can lead to the formation of a nonmagnetic Kondo--singlet state, for which an enhanced Pauli--like magnetic susceptibility results from the presence of a narrow peak in the quasiparticle DOS near the Fermi energy due to the Abrikosov--Suhl resonance.\cite{Lawrence, Evans} 

Since there are two distinct Ce sites in Ce$_2$Rh$_3$Sn$_5$ (see Section~\ref{structure}), we attempted to fit the experimental magnetic susceptibility in a broad temperature range as:  
\begin{equation}
\chi(T)= \frac{C}{T - \theta} + \chi_{\mathrm{Ce2}}(T),
\label{eq:chi-fit}
\end{equation}
with the first term describing Curie--Weiss behaviour anticipated for local moments of Ce$^{3+}$ ions from Ce1 sublattice and the second term $\chi_{\mathrm{Ce2}}$($T$) included to account for magnetic susceptibility of Ce2 ions calculated based on \mbox{$\chi$($T/T_{\mathrm{0}}$)/$\chi$(0)} curve obtained by Rajan\cite{Rajan} for trivalent Ce impurities ($j=5/2$) embedded in a see of conduction electrons. Here, $T_{\mathrm{0}}$ stands for a characteristic temperature that reflects the strength of coupling between the Ce~4$f$ and conduction band states, and $\chi$(0) denotes the Pauli--like magnetic susceptibility of a nonmagnetic Kondo--singlet state at zero temperature. A very good fit to the experimental $\chi$($T$) data at temperatures between $\sim$10~K and 400~K was achieved assuming \mbox{$T_0\approx 280$~K}, \mbox{$\chi(0)\approx 0.003$~emu/mol}, the paramagnetic Weiss temperature $\theta\approx-6$~K and the Curie constant $C$~=~0.48~emu~K/mol-Ce (see Fig.~\ref{fig:Fig6}).

The negative Weiss temperature of -6~K is comparable to the magnetic ordering temperature and concurs with the dominance of weak AFM exchange interactions between local magnetic moments of Ce ions from the Ce1 sublattice. The estimated $C$ value corresponds to the fluctuating moment of 1.96~$\mu_\mathrm{B}$ per Ce1, which is somewhat smaller than the magnetic moment of 2.54~$\mu_{\mathrm{B}}$ expected for free Ce$^{3+}$ ions.  
We note that the local environment of Ce1 ions in the crystal lattice of Ce$_2$Rh$_3$Sn$_5$ is very similar to that in CeRh$_2$Sn$_4$,\cite{moje124}, an antiferromagnet with well localized Ce$^{3+}$ moments. 
Since for CeRh$_2$Sn$_4$ the effective fluctuating moment decreases at temperatures below $\sim$150~K due to a thermal depopulation of excited crystal field levels from the $j=5/2$ multiplet of Ce$^{3+}$ ions,\cite{moje124} we suspect that the effective moment derived from the fit of the $\chi$($T$) for Ce$_2$Rh$_3$Sn$_5$ using Eq.~\ref{eq:chi-fit} may be influenced by crystal electric field (CEF) effect on the magnetic Ce1 ions.

Since Eq.~\ref{eq:chi-fit} does not consider CEF effects, the reliability of the performed fit is rather limited, especially regarding the detailed temperature dependence of the $\chi_{\mathrm{Ce2}}$ term which is small as compared to the magnetic susceptibility originating from local magnetic moments of Ce1 ions in Ce$_2$Rh$_3$Sn$_5$. In addition, Eq.~\ref{eq:chi-fit} does not account for the influence of valence fluctuations revealed by the XAS measurements (see Section~\ref{XAS-spectra}) on the magnetic susceptibility due to Ce2 ions in Ce$_2$Rh$_3$Sn$_5$.
Nevertheless, the temperature of the maximum in the $\chi_{\mathrm{Ce2}}$($T$) derived from the fit of $\sim$80~K is close to the value of the ratio \mbox{$C$(Ce$^{3+}$)/(3$\chi$(0)) $\approx$ 89~K,} ($C$(Ce$^{3+}$) is the Curie constant for Ce$^{3+}$), as expected for IV systems with nearly trivalent Ce.\cite{Lawrence} 
Furthermore, the maximum in the $\chi_{\mathrm{Ce2}}(T)$ occurs in the same temperature range in which our XAS study revealed the fastest changes in Ce valence in Ce$_2$Rh$_3$Sn$_5$.
This finding agrees with the observation of Lawrence $et$ $al$.,\cite{Lawrence} that a single energy scale describes both charge and spin excitations associated with intermediate--valent Ce ions in systems with Ce valence close to 3+. 
Finally, it is worthwhile to note that the $\chi_{\mathrm{Ce2}}$($T$) term obtained from the fit closely resembles the magnetic susceptibility for CeRhSi$_2$, a nonmagnetic IV system with similar Ce valence, in which Ce ions occupy only one lattice site.\cite{235Kaczorowski}

Based on the Coqblin--Schrieffer model for orbitally degenerate Ce$^{3+}$ ions, the Kondo temperature $T_{\mathrm{K}}$ describing the rate of spin fluctuations arising from hybridization of Ce~4$f$ and conduction band states is related to the characteristic temperature $T_{\mathrm{0}}$ as follows: \mbox{$T_{\mathrm{K}}$=2$\pi T_{\mathrm{0}} W$($N$)/$N$,} where the degeneracy factor $N$=6 ($N$=2$j$+1) and the Wilson number $W$($N$=6)=0.6464.\cite{TK1, TK2} Therefore, the performed analysis of the magnetic susceptibility of Ce$_2$Rh$_3$Sn$_5$ gives an estimate of the Kondo temperature for Ce2 ions of 189~K.

Finally, from the $\chi$($T$) data we conclude that the low temperature magnetic properties of Ce$_2$Rh$_3$Sn$_5$ can be explained as a superposition of the enhanced Pauli paramagnetism (\mbox{$\chi(0)\approx 0.003$~emu/mol}) due to delocalized 4$f$ electrons of Ce2 ions and a magnetic order followed by local moment paramagnetic behaviour resulting from the presence of well localized magnetic moments of trivalent Ce1 ions. To further test this scenario, we performed specific heat measurements.

\subsection{Specific heat}
\label{specific-heat}

\begin{figure*}
\includegraphics[width=0.95\textwidth,angle=0]{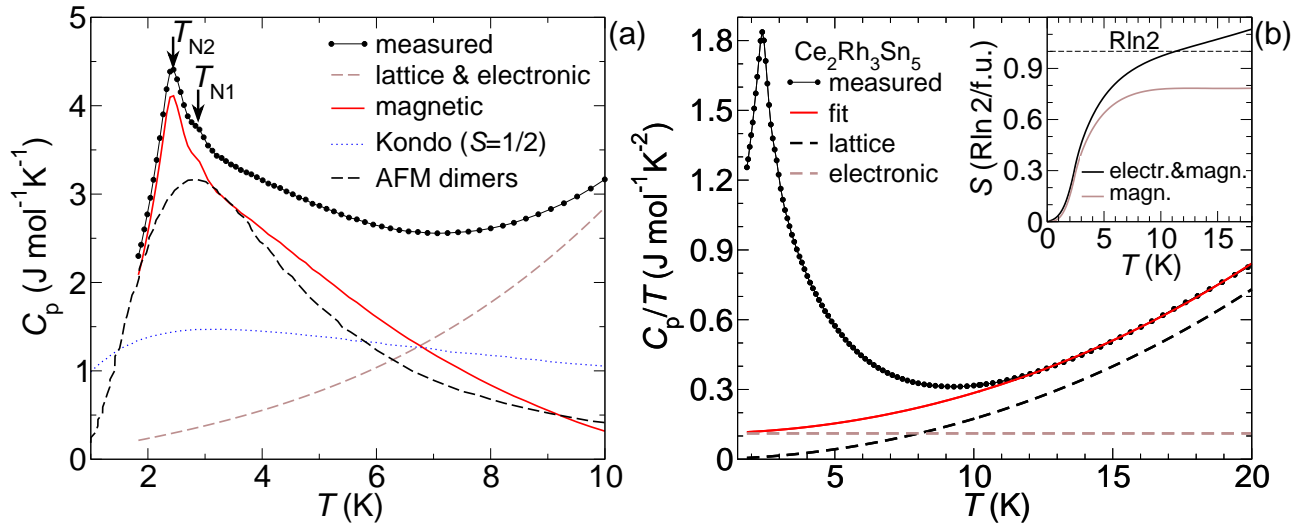}
\caption{(Color online) Low temperature specific heat of Ce$_2$Rh$_3$Sn$_5$ shown as $C_{\mathrm{p}}$($T$) (panel a) and plotted in a conventional \mbox{$C_p$/$T$($T$)} representation (panel b). The experimental data (black circles) are shown together with results of fits based on Debye model as described in the text. The magnetic contribution (red solid line) as well as fits based on Kondo model for $S=1/2$\cite{Schotte} (blue dots) and simulation of the magnetic specific heat of AFM dimers (black dashed line) are also presented in panel a.   
The inset in panel b depicts the entropy per formula unit of Ce$_2$Rh$_3$Sn$_5$ calculated from magnetic specific heat (solid brown line) and based on the sum of the electronic and magnetic specific heat (black solid line). The dashed horizontal line indicates the value of Rln2 expected for one Ce ion a magnetic doublet ground state. 
\label{fig:Fig4}}
\end{figure*}  

Fig.~\ref{fig:Fig4} shows the low temperature specific heat data for Ce$_2$Rh$_3$Sn$_5$. A distinct anomaly in the $C_{\mathrm{p}}$($T$) curve at 2.4~K and a shoulder at 2.9~K are consistent with results of magnetic measurements (see Section~\ref{susceptibility}) and indicate that Ce$_2$Rh$_3$Sn$_5$ undergoes two magnetic transitions. 
However, the low temperature specific heat is dominated by a broad contribution underneath the magnetic peaks. 
To inspect its origin, we perform quantitative analysis of different contributions to the low temperature specific heat.

First we evaluate the lattice and electronic specific heat. To this end, we plotted the $C_\mathrm{p}$ data in a $C_\mathrm{p}$($T$)/$T$ representation (see Fig.~\ref{fig:Fig4}b) and applied the Debye model. At temperatures between $\sim$11~K and $\sim$20~K, the experimental data follows the dependence \mbox{$C_{\mathrm{p}}$/$T$=$\gamma$ + $\beta T^2$ + $\delta T^4$,}   
where the Sommerfeld coefficient \mbox{$\gamma\approx 110$~mJ/(mol~K$^2$)} accounts for the electronic part of the specific heat and the consecutive two terms with \mbox{$\beta\approx 1.66\times 10^{-3}$~mJ/(mol~K$^4$)} and \mbox{$\delta\approx 4.04\times 10^{-7}$~mJ/(mol~K$^6$)} represent the first two terms in the Taylor expansion for the lattice specific heat.  
Although the fit was performed on the specific heat data at rather high temperatures, we note that the upper limit of the temperature range used for this analysis is only of 25\% of the temperature at which $\chi_{\mathrm{Ce2}}(T)$ shows a maximum due to spin excitations from Kondo singlet ground state to $j$=5/2 multiplet for nonmagnetic Ce ions (see Section~\ref{susceptibility}). Further, the obtained $\beta$ value corresponds to the Debye temperature of 227~K, which is close to Debye temperatures reported by Patil~$et$~$al.$\cite{Patil235} for isostructural and isoelectronic compounds with other rare earth elements.

To check whether the estimated $\gamma$ value correlates with the enhancement of the zero--temperature magnetic susceptibility \mbox{$\chi(0)\approx 3\times$10$^{-3}$~emu/mol} (see Section~\ref{susceptibility}) and thus could be explained as due to itinerant $f$ electrons in the FL state, we calculate the Sommerfeld--Wilson ratio  \mbox{$R_{\mathrm{SW}}$=$\pi^2$k$_{\mathrm{B}}^2 \chi$(0)/(3$\mu_{\mathrm{eff}}^2 \gamma$)} using the free--ion value of the effective fluctuating moment $\mu_{\mathrm{eff}}$ of $2.54 \mu_{\mathrm{B}}$ per Ce$^{\mathrm{3+}}$ ion. The resulting $R_{\mathrm{SW}}\approx 1$ is typical for HF systems in which strong hybridization between 4$f$ and conduction band states leads to the delocalization of the 4$f$ electrons, giving rise to an enhancement of the effective mass at low temperatures.\cite{VF-SW1}

The magnetic specific heat \mbox{$C_m$($T$)} was calculated by subtracting from the total $C_p$($T$) the estimated lattice and electronic contributions (see Fig.~\ref{fig:Fig4}a). The \mbox{$C_m$($T$)/$T$} was extrapolated to $T=0$ and then integrated with respect to temperature to give an estimate for the low temperature magnetic entropy, $S_{\mathrm{m}}$($T$).
As shown in the inset of Fig.~\ref{fig:Fig4}b, the $S_{\mathrm{m}}$($T$) saturates at the value of 0.8~Rln2 per formula unit. Even with the electronic specific heat term $\gamma T$ included in the estimate, the magnetic entropy is only of Rln2 per formula unit containing two Ce ions. This result implies that only half of Ce ions in Ce$_2$Rh$_3$Sn$_5$ is in a CEF doublet ground state and contributes to the low temperature magnetic ordering.

The magnetic entropy saturates well above the magnetic transitions only, at $T\approx10$~K (see inset of Fig.~\ref{fig:Fig4}b). Importantly, the value of the $S_{\mathrm{m}}$ recovered at $T_{\mathrm{N2}}$=2.9~K is only of 0.40~Rln2 per formula unit. Such a strong reduction of the magnetic entropy can be attributed to the Kondo effect on magnetic Ce ions and/or short range magnetic correlations that develop above magnetic ordering temperature.
Assuming that the effect results solely from partial quenching of Ce 4$f$--derived magnetic moments of Ce$^{3+}$ ions by Kondo effect, the single--ion Kondo temperature for magnetic Ce ions estimated based on model calculations of Yashima~$et$~$al.$\cite{Yashima} for $S=1/2$ Kondo impurity should be of 7~K. 
However, attempts to describe the broad contribution to the magnetic specific heat based on the Kondo model\cite{Schotte} do not give satisfactory results (see Fig.~\ref{fig:Fig4}a).
Instead, the shape of the \mbox{$C_m$($T$)} resembles magnetic anomalies observed for low dimensional systems.\cite{LD1, LD2} As an example, Fig.~\ref{fig:Fig4}a presents magnetic specific heat simulated for a system of AFM dimers with $S=1/2$ assuming that exchange coupling inside the dimers $J/k_{\mathrm{B}}\approx 5.7$~K and 75\% of the magnetic Ce ions are involved in the short range order. Close similarity between the calculated curve and the magnetic specific heat hints at a dominance of an AFM coupling between local moments of nearest neighbouring Ce$^{3+}$ ions. On the other hand, a layered crystal structure of Ce$_2$Rh$_3$Sn$_5$ composed of B blocks with trivalent Ce1 ions separated by A fragments containing nonmagnetic Ce2 ions (see Section~\ref{structure}) may facilitate quasi--two--dimensional magnetic correlations. Further studies including neutron diffraction and inelastic scattering measurements are needed to inspect in detail the low temperature magnetic order and magnetic excitations in Ce$_2$Rh$_3$Sn$_5$.

\subsection{Electrical resistivity}
\label{resistivity}

\begin{figure}
\includegraphics[width=0.46\textwidth,angle=0]{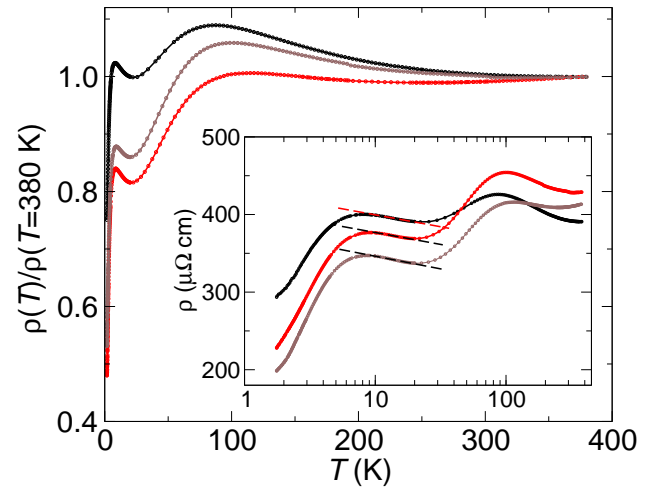}
\caption{\label{fig:Fig5} (Color online) The electrical resistivity of Ce$_2$Rh$_3$Sn$_5$ measured on three different polycrystalline blocks normalized to the value of the resistivity at 380~K. The inset shows the $\rho$($T$) data plotted as a function of temperature on a logarithmic scale.}
\end{figure}

Fig.~\ref{fig:Fig5} shows the temperature dependence of the electrical resistivity measured on three polycrystalline blocks of Ce$_2$Rh$_3$Sn$_5$. The overall shape of the $\rho$($T$) curves in conjunction with the values of the resistivity of \mbox{$\sim$200--300 $\mu \Omega$ cm} at lowest temperatures indicate a metallic behaviour of Ce$_2$Rh$_3$Sn$_5$, as expected based on the thermodynamic and spectroscopic data, and in agreement with previous report.\cite{Patil235}

For all the investigated specimens, the electrical resistivity shows broad maxima at $T\sim100$~K. At similar temperatures the XAS study revealed the fastest changes in Ce valence (see Section~\ref{XAS-spectra}) and the $\chi_{\mathrm{Ce2}}(T)$ shows maximum due to excitations from Kondo singlet ground state to $j$=5/2 multiplet (see Section~\ref{susceptibility}). Therefore, we explain the strong scattering in this temperature range as due to both charge and spin fluctuations associated with intermediate--valent Ce ions. Similar pronounced contributions to $\rho$($T$) were observed for many Ce--based IV systems.\cite{Lawrence, 235Kaczorowski} 

At lower temperatures, the $\rho$($T$) increases with decreasing temperature in a logarithmic fashion, as indicated in the inset of Fig.~\ref{fig:Fig5}. Such logarithmic upturns signify the dominance of the -ln$T$ term resulting from incoherent Kondo scattering of conduction electrons on magnetic Ce ions. 

Finally, the $\rho$($T$) curves pass through maxima at temperatures of 8~K and fall rapidly with decreasing temperature. Such a temperature dependence of the resistivity is a typical manifestation of an onset of coherence between Kondo scattering centres arranged periodically in a crystal lattice.\cite{Kondo-opor} The development of coherence governs the $\rho$($T$) at temperatures close to the magnetic transitions, which may explain the lack of evidence for the magnetic transitions in the electrical resistivity curves.

\subsection{Fixed spin moment calculations}
\label{FSMsection}

\begin{figure} 
\includegraphics[width=0.48\textwidth,angle=0]{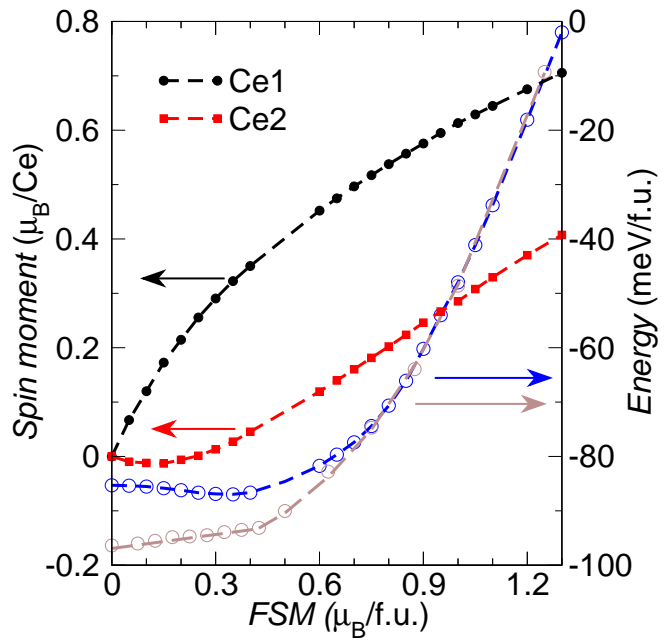}
\caption{\label{fig:Fig8} (Color online) The total energy--versus--FSM curves derived from FSM calculations for Ce$_2$Rh$_3$Sn$_5$ within the LSDA approximation assuming either the experimental crystal structure (empty beige circles) or the LDA crystal lattice (filled blue circles). The FSM values represent the total magnetic spin moments per formula unit. The total energy scale was shifted to allow for direct comparison of the curves obtained for the experimental and LDA crystal structures. 
The calculated values of Ce1 and Ce2 spin moments are shown as black circles and red squares, respectively. For the experimental crystal structure the symbols are empty, whereas for the LDA crystal lattice they are filled. The dashed lines are guided to the eye. The thin vertical dashed--dotted line separates FSM solutions with negligible spin polarization on Ce2 sites (on the left side) from those with Ce2 spin moment of at least 0.1~$\mu_\mathrm{B}$ (on the right side).}
\end{figure}

To get insight into magnetic properties of Ce$_2$Rh$_3$Sn$_5$ from first principles, we performed a series of electronic band structure calculations within the LSDA approximation using the fixed spin moment (FSM) method.\cite{FSM} This computational technique allows to fix the total magnetic spin moment of a system and adds this constraint to the DFT treatment.
 
Fig.~\ref{fig:Fig8} shows the total energies computed for different values of the FSM assuming the experimental crystal structure of Ce$_2$Rh$_3$Sn$_5$. The calculations indicate that the nonmagnetic state has the lowest total energy, but there is a flat region in the total energy--versus--FSM curve ranging from 0 to about \mbox{0.4 $\mu_\mathrm{B}$/f.u.}, suggesting a proximity of a magnetic state.  

Since calculated magnetic properties are often sensitive to atomic coordinates, we performed their computational relaxation based on atomic forces\cite{forces} to find theoretical equilibrium atomic positions for the LDA approximation. The resulting internal coordinates for which total force on each atom is smaller than 5~mRy/a.u. are listed in Table~\ref{tab:Table2}.  
For this relaxed crystal structure, the LSDA calculations with initial spin--polarization converged to a magnetic state, in agreement with the results of thermodynamic measurements (see Sections~\ref{susceptibility}, \ref{specific-heat}). FSM calculations show that there is a shallow magnetic minimum in the energy--versus--FSM curve at the FSM value of \mbox{0.35 $\mu_\mathrm{B}$/f.u.}, that has now slightly lower total energy than the nonmagnetic solution (\mbox{$\Delta E\approx$ 1.7 meV/f.u;}  see Fig.~\ref{fig:Fig8}). Remarkably, the magnetic moment of \mbox{0.35 $\mu_\mathrm{B}$/f.u.} is carried basically only by Ce1 atoms, whereas Ce2 atoms as well as all Rh and Sn atoms in Ce$_2$Rh$_3$Sn$_5$ stay almost unpolarized (spin moments below 0.05~$\mu_\mathrm{B}$ per atom). 

Obviously the detailed shape of the calculated energy--versus--FSM curve should depend not only on the atomic positional parameters but also on the lattice parameters. Furthermore, the symmetry lowering required by various possible spin arrangements may also affect slightly the calculated total energy--versus--FSM curves. 
Nevertheless, the performed calculations for the experimental and the relaxed crystal structures indicate there is a notable energy cost associated with inducing spin polarization on Ce2 sites in Ce$_2$Rh$_3$Sn$_5$ (Fig.~\ref{fig:Fig8}). The minimum difference in total energy between FSM calculations giving magnetic spin moment for Ce2 below $\sim$0.05~$\mu_\mathrm{B}$ (FSM~$\textless$0.45~$\mu_\mathrm{B}$/f.u.; region on the left side of the vertical line on Fig.~\ref{fig:Fig8}) and calculations resulting in spin polarization on Ce2 sites of at least 0.1~$\mu_\mathrm{B}$ per Ce2 (region on the right side of the vertical line on Fig.~\ref{fig:Fig8}) is of 3.5~meV/f.u. Furthermore, for FSM calculations resulting in magnetic spin moment for Ce2 of 0.1~$\mu_\mathrm{B}$, the calculated moments on Ce1 sites are of 0.35--0.45~$\mu_\mathrm{B}$. 
Therefore, we conclude that Ce1 ions should bear magnetic moments and contribute to the low temperature magnetic ordering in Ce$_2$Rh$_3$Sn$_5$, whereas the Ce2 ions should remain nonmagnetic.

\subsection{Valence band}
\label{valence}

\begin{figure}
\includegraphics[width=0.46\textwidth,angle=0]{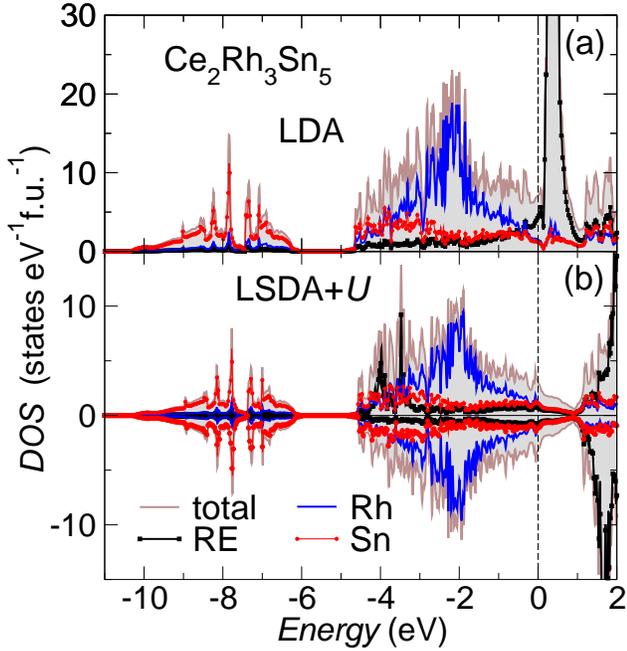}
\caption{\label{fig:Fig9} (Color online) The total and atomic resolved DOSs for Ce$_2$Rh$_3$Sn$_5$ calculated within the LDA approximation (panel~a) and using the LSDA+$U$ approach for the Ce~4$f$ shell (panels~b,c) with \mbox{$U_{\mathrm{eff}}$ = 6 eV\cite{typoweU}} (panel b; thin lines with dots in panel c) or \mbox{$U_{\mathrm{eff}}$ = 3 eV} (thick solid lines in panel c).}
\end{figure} 

\begin{figure}
\includegraphics[width=0.45\textwidth,angle=0]{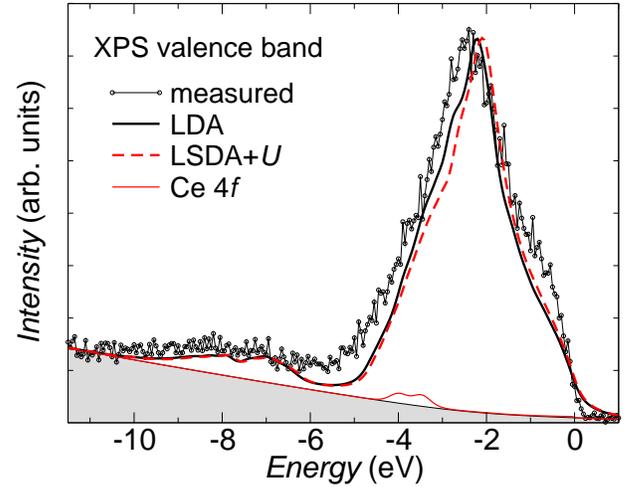}
\caption{\label{fig:Fig10} (Color online) XPS valence band spectrum of Ce$_2$Rh$_3$Sn$_5$ (thin black solid line with dots) in comparison with theoretical spectra calculated based on partial DOSs obtained within the LDA approximation (thick black line) and using the LSDA+$U$ ($U_{\mathrm{eff}}$~=~6~eV)\cite{typoweU} approach for the Ce~4$f$ shell (thick dashed red line). The thin red (grey) solid line shows the sum of partial Ce1~4$f$ and Ce2~4$f$ contributions. The grey field represents background estimated using Tougaard algorithm\cite{Tou}.}
\end{figure}

Fig.~\ref{fig:Fig9} shows the total and partial atomic resolved DOSs for Ce$_2$Rh$_3$Sn$_5$ calculated for the experimental crystal structure within the LDA approximation and using the LSDA+U approach to account for the strong Coulomb interaction within Ce~4$f$ shell. The valence band of Ce$_2$Rh$_3$Sn$_5$ consists of two parts separated by the gap of 1.5~eV. The structure at binding energies ranging from 5 to 8~eV originates primarily from 5$s$ states of Sn atoms. The main part of the valence DOS is dominated by hybridized Rh~4$d$ and Sn~5$p$ states. 

Within the LDA approximation, the Ce~4$f$ states of both Ce1 and Ce2 form narrow bands with the centre of gravity above the Fermi level (see Fig.~\ref{fig:Fig9}a). 
The calculated number of 4$f$ electrons for Ce1 ions equals 1.01. Thus, the electronic configuration of Ce1 is very close to 4$f^1$ expected for Ce$^{3+}$. 
For Ce2 ions, however, the number of occupied 4$f$ states is of 0.92. Such a reduced occupation of the Ce2 4$f$ shell suggests a small deviation from the trivalent configuration, in line with a valence slightly larger than 3+ revealed by our XAS measurements (see Section \ref{XAS-spectra}).  

Since the LDA method underestimates the Coulomb repulsion of electrons within narrow bands, we performed also additional band structure calculations using the LSDA+$U$ approach to account for stronger Coulomb interaction within Ce 4$f$ shell. Inclusion of the Hubbard-like interaction terms to the XC potential results in a shift of the occupied Ce~4$f$ bands toward higher binding energies and of the unoccupied 4$f$ states above the Fermi level for Ce ions in both lattice sites. As an example, Fig.~\ref{fig:Fig9}c shows the partial DOSs for Ce1 and Ce2 calculated using the LSDA+$U$ method assuming $U_{\mathrm{eff}}$=3~eV or $U_{\mathrm{eff}}$=6~eV. 

Our calculations show that for Ce1 the application of the $U_{\mathrm{eff}}$ parameter larger than $\sim$2~eV yields the qualitatively correct physical picture of Ce$^{3+}$ with magnetic spin moments of 1~$\mu_\mathrm{B}$/Ce and the occupied 4$f$ states forming narrow bands well below the $E_{\mathrm{F}}$. 
In case of Ce2, however, somewhat larger values of the $U_{\mathrm{eff}}$ of at least 3~eV were required to suppresses the hybridization of the Ce2 4$f$ and valence band states near the Fermi level and to move the 4$f$ bands away from the Fermi energy. We note that for $U_{\mathrm{eff}}\gtrsim$3~eV the binding energies of the occupied 4$f$ states are smaller by about 0.5~eV for Ce2 than for Ce1, independent of the exact value of the $U_{\mathrm{eff}}$ parameter (see Fig.~\ref{fig:Fig9}c).
Furthermore, the changes in the Ce2~4$f$ bands induced by application of the LSDA+$U$ approach lead to a narrowing of the main part of the valence band and result in a shift of 4$d$ states of Rh2 and Rh3 towards lower binding energies, whereas the strength of correlation effects within the 4$f$ shell of Ce1 ions has almost no impact on the other valence band states (not shown). Thus, the computational study gives an indirect evidence for the importance of the hybridization between Ce2~4$f$ and conduction band states in Ce$_2$Rh$_3$Sn$_5$.

To shed light on the effective mass enhancement in Ce$_2$Rh$_3$Sn$_5$, we estimate bare values of the Sommerfeld coefficient \mbox{$\gamma_b =$($\pi^2$/3)$k_{\mathrm{B}}$DOS($E_{\mathrm{F}})$} using the DOS($E_{\mathrm{F}})$ derived from our calculations of \mbox{6--7 st./(eV f.u.)} (Fig.~\ref{fig:Fig9}).  
The obtained $\gamma_b$ values of \mbox{15--17 mJ/(mol K$^2$)} are about seven times smaller than the experimental Sommerfeld coefficient in the paramagnetic region of \mbox{110 mJ/(mol K$^2$)} (see Section \ref{specific-heat}). Different effects such as phonon--electron coupling or low--lying magnetic excitations can enlarge the bare $\gamma_b$ value. However, such a strong enhancement of the Sommerfeld coefficient suggests that there is a notable renormalization of the effective quasiparticle masses due to dynamic electron--electron correlations. Since the applied computational methods (LDA and LSDA+$U$) are static mean--field approximations, by definition they neglect all dynamic correlation effects such as renormalization of electronic bands and the formation of Abrikosov--Suhl resonance. Therefore, we conclude that none of our calculations gives a proper description of the electronic structure of Ce$_2$Rh$_3$Sn$_5$ in a region close to the Fermi level. Further theoretical study based on dynamical mean--field theory (DMFT) is needed to inspect in detail the shape of the DOS in the vicinity of the Fermi energy for Ce$_2$Rh$_3$Sn$_5$.

In order to gain experimental insight into the valence band of Ce$_2$Rh$_3$Sn$_5$, we performed photoemission measurements. Fig.~\ref{fig:Fig10} shows the XPS valence band spectrum for Ce$_2$Rh$_3$Sn$_5$. 
To facilitate its comparison with the band structure results, the theoretical XPS spectra were estimated based on the calculated partial DOSs as described in Section~\ref{computational}. The commonly used Tougaard--type\cite{Tou} background (grey field in Fig.~\ref{fig:Fig10}) was added to the calculated data to account for the presence of secondary electrons during photoemission processes. The exemplary results obtained based on DOSs calculated within the LDA approximation and using the LSDA+$U$ ($U_{\mathrm{eff}}$~=~6~eV) approach are presented in Fig.~\ref{fig:Fig10}.

The calculated curves reflect all the essential features present in the measured XPS valence band spectrum, including the energy gap between the Sn~4$s$--like shallow core states and the remainder of the valence band. Thus, the experiment confirms the reliability of our computational results. Although the LDA approximation seems to give a slightly better match regarding energy positions of the main features in the XPS valence band spectrum than the LSDA+U approach, the limited energy resolution of the performed XPS experiment prevents more detailed comparisons. Further, the Ce~4$f$ contributions to the XPS valence band spectrum are very small as compared to the contributions from the other valence band states (Fig.~\ref{fig:Fig10}). Therefore, the performed experiment also cannot give direct information about the localization of the Ce~4$f$ states in the conduction band of Ce$_2$Rh$_3$Sn$_5$.

\section{Conclusions and summary} 
\label{Conclusions}

The crystal structure of Ce$_2$Rh$_3$Sn$_5$ was refined from single crystal XRD data. Analysis of interatomic distances and coordination of Ce atoms occupying two distinct crystallographic sites suggests that Ce1 species are in a trivalent state (4$f^1$), whereas the valence of Ce2 ions should be larger than +3 due to extraordinarily short Ce2--Rh contacts.
Electronic band structure calculations provide further indication for a trivalent state of Ce1 ions. By contrast, for Ce2 ions a reduced occupation of 4$f$ states to $\sim$0.92 suggests a small deviation from the trivalent electronic configuration. Importantly, XAS measurements revealed an IV behaviour. Assuming that Ce1 atoms are trivalent in the entire investigated temperature range, the valence of Ce2 species derived from the deconvolution of Ce~$L_{\mathrm{III}}$~XAS spectra varies between +3.18 at 6~K and +3.08 at $\sim$300~K. 

Thermodynamic measurements revealed a magnetic ground state for Ce$_2$Rh$_3$Sn$_5$. Both specific heat and magnetization data show two magnetic transitions at $T_{\mathrm{N1}}\approx$~2.9~K and $T_{\mathrm{N2}}\approx$~2.4~K. The low temperature magnetic entropy of Rln2 per formula unit implies, however, that only half of Ce ions in Ce$_2$Rh$_3$Sn$_5$ participates in the magnetic ordering. Electronic band structure calculations indicate that only Ce1 ions exhibit spin polarization in Ce$_2$Rh$_3$Sn$_5$. For Ce2 ions, the computational results in conjunction with the XPS spectra point to the importance of the hybridization of the 4$f$ and conduction band states. 
This hybridization is crucial for the formation of a nonmagnetic Kondo--singlet state characterized by an enhanced Pauli--like magnetic susceptibility \mbox{$\chi(0)\approx 0.003$~emu/mol} and a large electronic specific heat \mbox{$\gamma\approx 110$~mJ/(mol~K$^2$)}, which is about seven times larger than the bare $\gamma_b$ value derived from the calculated single--particle DOS($E_{\mathrm{F}}$). Such a strong augmentation of the Sommerfeld coefficient supports the important role of electron--electron correlation effects, whereas the Sommerfeld--Wilson ratio $R_{\mathrm{SW}}\approx 1$ provides a further indication that both $\chi$(0) and $\gamma$ are similarly enhanced due to heavy quasiparticles in the FL state.

With increasing temperature, the strong Pauli--like behaviour in the magnetic susceptibility is going over to a local moment paramagnetism. Finally, near room temperature Ce$_2$Rh$_3$Sn$_5$ behaves as an ordinary paramagnet with nearly full moment expected for all the Ce ions being in a trivalent state and a small positive \mbox{$\chi_{\mathrm{0-HT}}\approx 8\times10^{-5}$~emu/mol} which, after correcting for core level diamagnetism and Landau diamagnetism, is comparable to the Pauli susceptibility expected based on the DOS($E_{\mathrm{F}}$) derived from our electronic structure calculations.

To summarize, our combined experimental and theoretical study indicates that Ce$_2$Rh$_3$Sn$_5$ is a Kondo lattice system in which at low temperatures a magnetic order due to trivalent Ce1 ions coexists with an intermediate--valence behaviour of Ce2 ions with a nonmagnetic Kondo--singlet ground state. Further study is needed to elucidate in detail the electronic band structure of Ce$_2$Rh$_3$Sn$_5$ near the Fermi level where dynamical many--body effects determine the shape of quasiparticle DOS and to sheed more light on the magnetic ground state.

\section{Acknowledgments}

The authors thank Dr Christoph Geibel from Max-Planck Institute for Chemical Physics of Solids for fruitful discussions. The authors are grateful to Dr Jerzy Kubacki from University of Silesia for his kind help with XPS measurements and to Dr E. Welter and Dr D. Zajac from HASYLAB for their helpful assistance during XAS measurements on beamline C of Hasylab, Desy. M. G. would like to acknowledge financial support from the Max--Planck Society and the DAAD foundation through research fellowships.

\end{document}


\preprint{APS/123-QED}            

\title{Coexistence of magnetic order and valence fluctuations in the Kondo lattice system Ce$_2$Rh$_3$Sn$_5$.}

\author{M. B. Gam\.za$^{1,2,3}$ \email{MGamza@uclan.ac.uk}, R. Gumeniuk$^{2,4}$, U. Burkhardt$^{2}$, W. Schnelle$^{2}$, H. Rosner$^{2}$, A. Leithe-Jasper$^{2}$, and A.~\'{S}lebarski$^{3,5}$}

\affiliation{$^{1}$Jeremiah Horrocks Institute for Mathematics, Physics and Astrophysics, University of Central Lancashire, Preston PR1 2HE, UK}
\affiliation{$^{2}$Max-Planck Institute for Chemical Physics of Solids, D-01187 Dresden, Germany}
\affiliation{$^{3}$Institute of Physics, University of Silesia, 40-007 Katowice, Poland}
\affiliation{$^{4}$Institute of Experimental Physics, TU Bergakademie Freiberg, 09596 Freiberg, Germany}
\affiliation{$^{5}$Centre for Advanced Materials and Smart Structures, Polish Academy of Sciences, 50-950 Wroc\l aw, Poland}

\maketitle

\section{Supplementary Information}

\subsection{Metallographic study}

The microstructures of the Ce$_2$Rh$_3$Sn$_5$ sample was inspected optically (Zeiss Axioplan~2) and with a scanning electron microscope (Philips~XL~30). For the metallographic examination, a piece of about 3~mm diameter was cut from the annealed sample and embedded in conductive resin. Grinding was performed on abrasive papers (500- and 1000-grid silicon carbide). Polishing was done using slurries of 9, 3 and 1/4 $\mu$m diamond powder in alcohol-based lubricants. 

\begin{figure}
\includegraphics[width=0.5\textwidth,angle=-90]{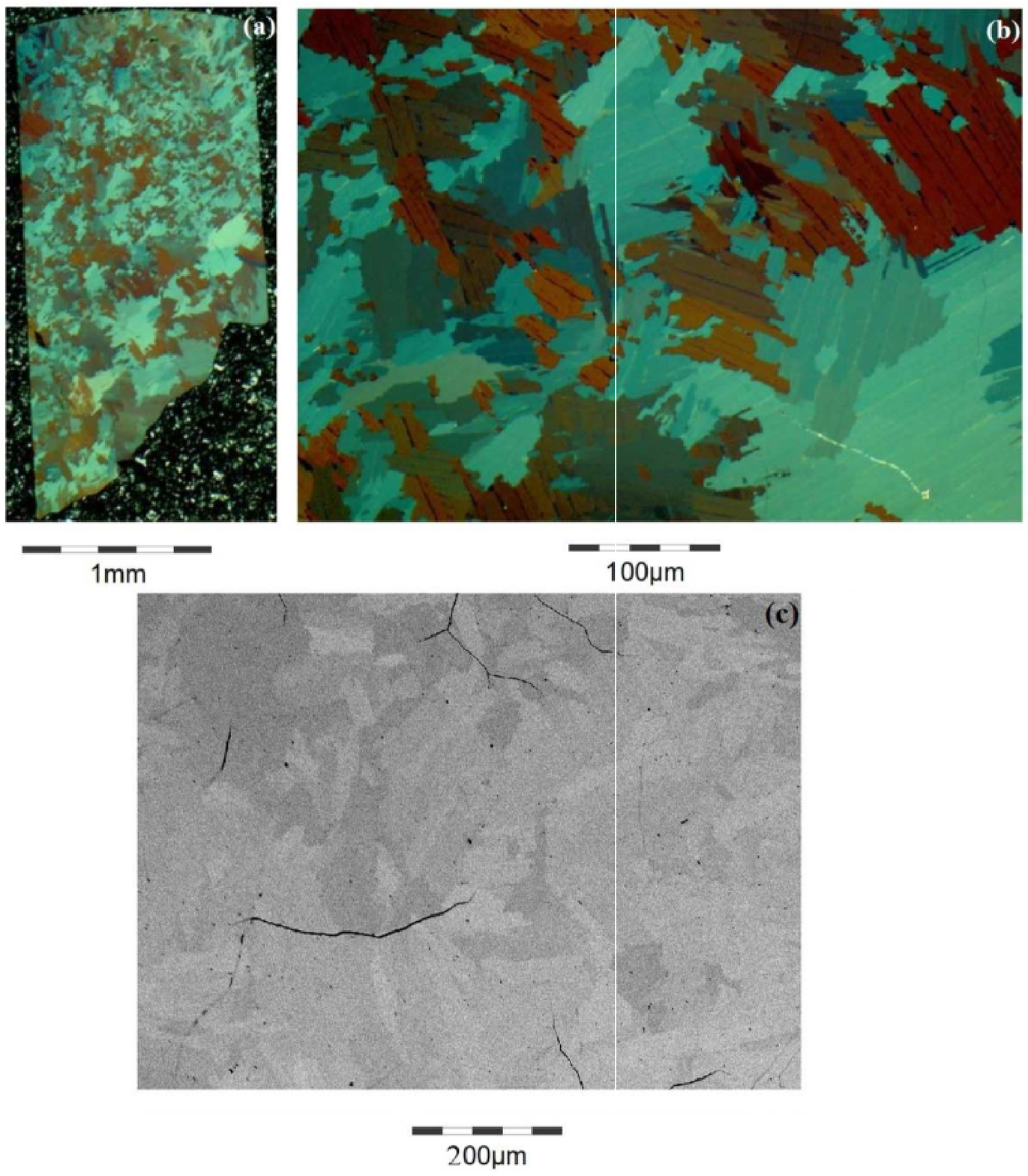}
\caption{\label{fig:Fig1} Metallographic cross--section of an annealed (800$^{\circ}$C/14 days) polycrystalline sample of Ce$_2$Rh$_3$Sn$_5$ showing orientation dependent reflectivity of the grains in optical (polarized light mode, panels a, b) and scanning electron (back scattered electron mode, panel c) micrographs.}
\end{figure}

Images of the sample surface obtained using the optical microscope with polarized light suggest that the sample consists of elongated grains with typical lengths of between 10~$\mu$m and 100~$\mu$m oriented randomly in various directions (see panels a and b of Fig.~\ref{fig:Fig1}). Scanning electron microscopy investigations indicate that basically all the grains belong to the Ce$_2$Rh$_3$Sn$_5$ phase. Multiple images recorded from the same sample surface with a secondary electron detector and using a backscattered electron detector with magnifications ranging from 120$\times$ to 4000$\times$ did not detect impurity phases and shown only a uniform sample density.

\begin{figure} 
\includegraphics[width=0.48\textwidth,angle=0]{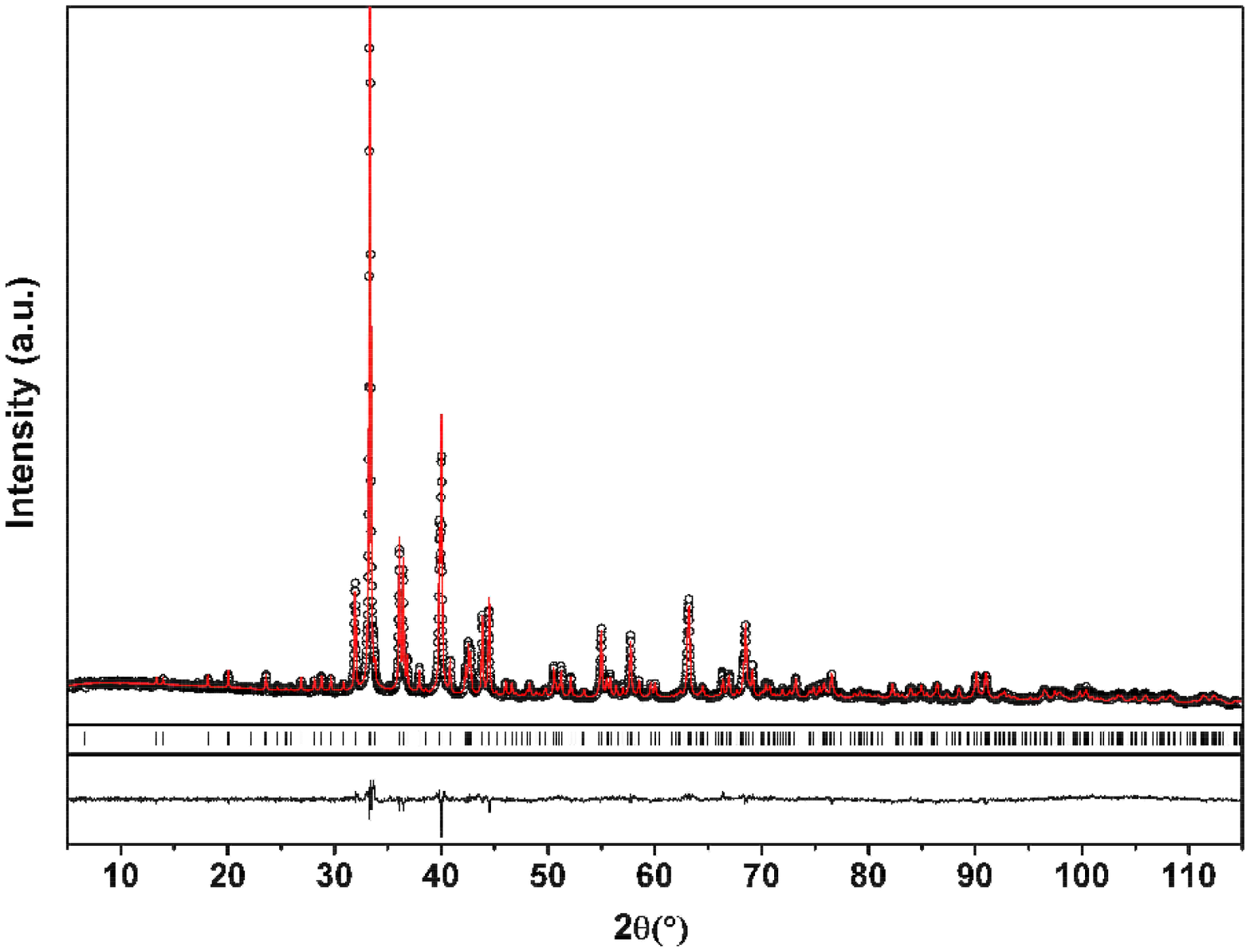}
\caption{\label{fig:Fig2} \small X-ray powder diffraction pattern for Ce$_2$Rh$_3$Sn$_5$ (black circles) and Rietveld refinement (red lines) that resulted in the following statistical factors: \mbox{$R$(all) = 6.92;} \mbox{$R_{\mathrm{w}}$(all) = 7.25;} \mbox{$R_{\mathrm{p}}$ = 5.17;} \mbox{$R_{\mathrm{wp}}$ = 6.59.}}
\end{figure}  

The chemical composition was investigated by means of wavelength dispersive X--ray spectroscopy (WDXS) using a CAMECA~SX100 electron microprobe equipped with a tungsten cathode. The local composition was determined by intensities of the X--ray lines Ce$L$ , Sn$L$  and Rh$L$  which were excited by an electron beam of 20~nA at 15~keV. The X--rays were focused by large monochromator crystals PET (Pentaerythritol, $d$~=~0.437~nm) on a gas flow proportional counter. The proportions of the three elements were determined with respect to the appropriate reference materials Rh, Sn and CeAl$_2$ and resulted in analytical totals of 100.1(2) wt.\% by using the PAP matrix correction model.\cite{PAP} Measurements on ten points on the sample surface gave the same results within expected experimental error bars (see Table~\ref{tab:Table1}). After averaging on the ten points, the following contents were obtained: Ce:~20.06(10)~at.\%, Rh:~29.95(19)~at.\%, Sn:~49.99(18)~at.\%. This composition corresponds to Ce$_{2.01(1)}$Rh$_{3.00(2)}$Sn$_{5.00(2)}$ and is in an excellent agreement with stoichiometric Ce$_2$Rh$_3$Sn$_5$.

\subsection{X-ray powder diffraction}

Phase analysis of the polycrystalline sample was carried out from X--ray powder diffraction patterns collected at room temperature on a HUBER imaging plate Guinier camera G670 using Cu~$K\alpha_1$ radiation in a 2$\theta$ range of 3--115~degrees with the expose time 6~$\times$~15~min. For powder XRD study, pieces of the sample were converted to a fine powder by grinding in a mortar with a small amount of acetone. 
Rietveld--type refinement of the powder pattern performed using Jana2006 program\cite{jana2006} gave atomic positional and displacement parameters consistent with results of our single crystal X-ray diffraction study.  Lattice parameters derived from the refinement are: \mbox{$a$ = 4.4992(1) \r{A}};  \mbox{$b$ = 26.4839(7) \r{A}}; \mbox{$c$ = 7.2160(2) \r{A}}. The sample was found to be nearly single phased. Apart from peaks originating from the Ce$_2$Rh$_3$Sn$_5$ phase, there are only few additional very slight features in the diffraction pattern,  which we assign to a small amount of an unidentified minority phase (see Fig.~\ref{fig:Fig2}).

\begin{table}
\caption{\label{tab:Table1} Normalized atomic concentrations of Ce, Rh and Sn measured by WDXS method on ten points on the Ce$_2$Rh$_3$Sn$_5$ sample surface.}
\begin{tabular}{l|l|l|l}
Nr.  & Ce (at.\%) & Rh (at.\%)   & Sn (at.\%) \\ \hline
1    & 20.03      & 30.26        & 49.71      \\   
2    & 20.06      & 29.79        & 50.14      \\
3    & 20.10      & 30.00        & 49.90      \\
4    & 20.01      & 29.81        & 50.18      \\
5    & 20.12      & 30.03        & 49.86      \\
6    & 20.08      & 29.76        & 50.16     \\
7    & 19.99      & 30.15        & 49.86     \\
8    & 19.94      & 29.81        & 50.25     \\
9    & 19.95      & 30.16        & 49.89     \\
10   & 20.28      & 29.77        & 49.95     \\  \hline
 \end{tabular}
\end{table}